\documentclass[12pt]{iopart}

\usepackage{iopams}  
\usepackage{graphicx}
\usepackage{epsfig} 
\begin{document}

\title{$p$-Process simulations with a modified reaction library}

\author{I Dillmann$^{1,2}$, T Rauscher$^2$, M Heil$^{1,3}$, F K{\"a}ppeler$^1$, W Rapp$^4$, and F-K Thielemann$^2$}
\address{$^1$ Institut f{\"u}r Kernphysik, Forschungszentrum Karlsruhe, Postfach 3640, D-76021 Karlsruhe} \ead{iris.dillmann@ik.fzk.de}
\address{$^2$ Departement Physik und Astronomie, Universit{\"a}t Basel, Klingelbergstrasse 82, CH-4056 Basel}
\address{$^3$ Gesellschaft f{\"u}r Schwerionenforschung mbH, Planckstrasse 1, D-64291 Darmstadt }
\address{$^4$ Westinghouse Electric Germany GmbH, Dudenstrasse 44, D-68167 Mannheim}

\begin{abstract}
We have performed $p$-process simulations with the most recent stellar $(n,\gamma)$ cross sections from the "Karlsruhe Astrophysical Database of Nucleosynthesis in Stars" project (version v0.2, http://nuclear-astrophysics.fzk.de/kadonis). The simulations were carried out with a parametrized supernova type II shock front model (``$\gamma$ process'') of a 25 solar mass star and compared to recently published results. A decrease in the normalized overproduction factor could be attributed to lower cross sections of a significant fraction of seed nuclei located in the Bi and Pb region around the $N$=126 shell closure.

\end{abstract}
\submitto{\JPG}
\maketitle

\section{The ``$p$ processes''}
A ``$p$ process'' was postulated to produce 35 stable but rare isotopes between $^{74}$Se and $^{196}$Hg on the proton-rich side of the valley of stability. Unlike the remaining 99\% of the heavy elements beyond iron these isotopes cannot be created by (slow or rapid) neutron captures \cite{bbfh57}, and their solar and isotopic abundances are 1-2 orders of magnitude lower than the respective $s$- and $r$-process nuclei \cite{AnG89,iupac}. However, so far it seems to be impossible to reproduce the solar abundances of all $p$ isotopes by one single process. In current understanding several (independently operating) processes seem to contribute.

The largest fraction of $p$ isotopes is created in the ``$\gamma$ process'' by sequences of photodissociations and
$\beta^+$ decays \cite{woho78,woho90,ray90}. This occurs in explosive O/Ne burning during SNII explosions and reproduces the solar abundances for the bulk of $p$ isotopes within a factor of $\approx$3 \cite{ray90,raar95}. The SN shock wave induces temperatures of 2-3 GK in the outer (C, Ne, O) layers, sufficient for triggering the required photodisintegrations. More massive stellar models (M$\geq$20~M$_\odot$) seem to reach the required temperatures for efficient photodisintegration already at the end of hydrostatic O/Ne burning \cite{RHH02}. 
The decrease in temperature after passage of the shock leads to a freeze-out via neutron captures and mainly $\beta^+$ decays, resulting in the typical $p$-process abundance pattern with maxima at $^{92}$Mo ($N$=50) and $^{144}$Sm ($N$=82).

However, the $\gamma$ process scenario suffers from a strong underproduction of the most abundant $p$ isotopes, $^{92,94}$Mo and $^{96,98}$Ru, due to lack of seed nuclei with $A>$90. For these missing abundances, alternative processes and sites have been proposed, either using strong neutrino fluxes in the deepest ejected layers of a SNII ($\nu p$ process \cite{FML06}), or rapid proton-captures in proton-rich, hot matter accreted on the surface of a neutron star ($rp$ process \cite{scha98,scha01}). A few $p$ nuclides may also be produced by neutrino-induced reactions during the $\gamma$-process. This "$\nu$ process" \cite{WHH90} was additionally introduced because the $\gamma$ process alone strongly underproduces the odd-odd isotopes $^{138}$La and $^{180m}$Ta. These two isotopes could be the result of excitation by neutrino scattering on pre-existing $s$-process seed nuclei, depending on the still uncertain underlying nuclear physics.

Modern, self-consistent studies of the $\gamma$-process have problems to synthesize $p$ nuclei in the regions $A<124$ and $150\leq A\leq 165$ \cite{RHH02}. It is not yet clear whether the observed underproductions are only due to a problem with astrophysical models or also with the nuclear physics input, i.e. the reaction rates used. Thus, the reduction of uncertainties in nuclear data is strictly necessary for a consistent understanding of the $p$ process. Experimental data can improve the situation in two ways, either by directly replacing predictions with measured cross sections in the relevant energy range or by testing the reliability of predictions at other energies when the relevant energy range is not experimentally accessible. In this context we have carried out $p$-process network calculations with a modified reaction library which uses the most recent experimental and semi-empirical $(n,\gamma)$ cross sections from the "Karlsruhe Astrophysical Database of Nucleosynthesis in Stars" project, KADoNiS v0.2 \cite{kado06}. This aims to be a step towards an improved reaction library for the $p$ process, containing more experimental data. However, it has to be kept in mind that the largest fraction of the $p$-process network contains proton-rich, unstable isotopes which are not accessible for cross section measurements with present experimental techniques. Hence there is no alternative to employing theoretical predictions for a large number of reactions. Typically, these come from Hauser-Feshbach statistical model calculations \cite{hafe52} performed with the codes NON-SMOKER \cite{rath00,rath01} or MOST \cite{most05}.


\section{$p$-process network calculations}
We studied the $p$ process in its manifestation as a $\gamma$ process. The network calculations were carried out with the
program "\textsc{pProSim}" \cite{RGW06}. The underlying network was originally based on a reaction library from Michigan State University for X-ray bursts which included only proton-rich isotopes up to Xenon. For $p$-process studies it was extended by merging it with a full reaction library from Basel university \cite{reaclib}. That reaction library was mainly based on NON-SMOKER predictions with only few experimental information for light nuclei. For the present calculations, the library was updated by inclusion of more than 350 experimental and semi-empirical stellar ($n,\gamma$) cross sections from the most recent version of the "Karlsruhe Astrophysical Database of Nucleosynthesis in Stars" (KADoNiS v0.2). Due to detailed balance this modification also affects the respective ($\gamma,n$) channels. 

The abundance evolution was tracked with a parameterized reaction network, based on a model of a
supernova type II explosion of a 25 M$_\odot$ star \cite{raar95}. Since the $p$-process
layers are located far outside the collapsing core, they only experience the explosion shock front passing through the O/Ne
burning zone and the subsequent temperature and density increase. Both, the seed abundances and the respective temperature and density profiles, were taken from \cite{raar95} and are not calculated self-consistently. The $p$-process zone in the simulation was subdivided into 14 single layers. This is consistent with what was used in \cite{RGW06} and thus the results can directly be compared.

The final $\gamma$-process abundances depend very sensitively on the choice of the initial seed abundance. This initial abundance
is produced by in-situ modification of the stellar material in various nucleosynthesis processes during stellar evolution.
This also means that the respective O/Ne layers can receive an abundance contribution from the weak $s$ process (core helium and shell carbon burning during the Red Giant phase of the massive star) in the mass
region up to $A$=90. The $s$-process component depends on the mass of the star and also on the neutron yield provided by the
$^{22}$Ne($\alpha,n$)$^{25}$Mg neutron source. If the $p$-process layers during the explosion are located within the convective
zones of the previous helium burning phases, the $s$-process abundance distribution in all layers can be assumed to be constant. Thus, it has to be emphasized that the results strongly depend on the adopted stellar model (e.g. Ref. \cite{RHH02} found significant $\gamma$-processing occurring already in late stages of hydrostatic burning). Additionally, stars with different masses will exhibit different $s$- and $p$-processing and true $p$-abundances can only be derived when employing models of galactical chemical evolution. Nevertheless, for better comparison with previous studies, here we implemented the same approach as used in \cite{ray90,raar95,RGW06} (see also Fig.~\ref{opf}).

\section{Results and Interpretation}
\begin{figure}[!htb]
\begin{center}
\includegraphics{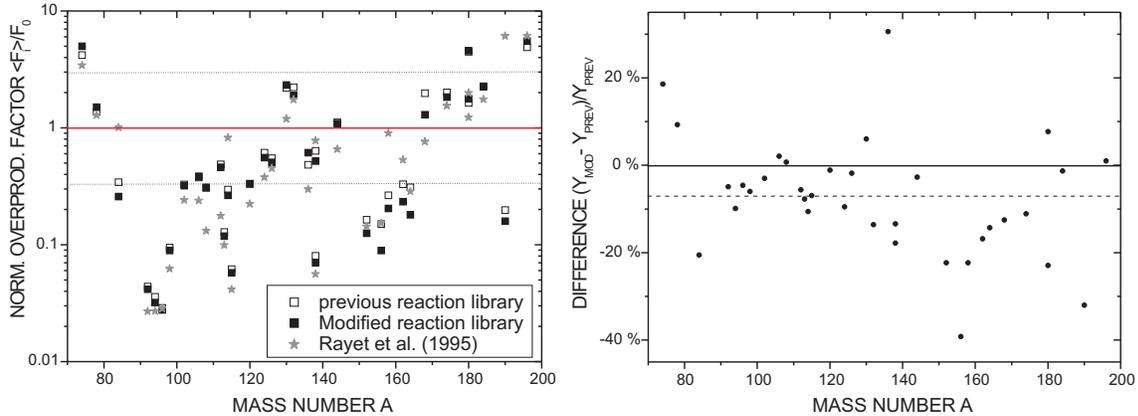}
\caption{Left: Normalized overproduction factors derived with the previous \cite{RGW06} (open squares)
and the modified (full squares) reaction library. Additionally the
values from a 25M$_\odot$ star model of Rayet et al. \cite{raar95}
are given for comparison (grey stars). A value equal to unity corresponds to the solar abundance. Right: Abundance difference for each $p$ isotope between the modified and the previous reaction library.}\label{opf}
\end{center}
\end{figure}

The results of our simulations are shown in Fig.~\ref{opf} as 'normalized overproduction factors' $<$$F_i$$>$/$F_0$ \cite{ID06}. This value gives the produced abundance relative to the solar abundances of Anders and Grevesse \cite{AnG89}. Ranges of variations of this factor for SN type II explosions with stellar masses 13~$M_\odot$$\leq M_{star}$$\leq$25~$M_\odot$ are published in Fig.~4 of Ref. \cite{raar95}. The factor $F_0$ is the so-called 'averaged overproduction factor' which is a measure for the overall enrichment of all $p$ isotopes. This value changed only by -0.6\% from $F_0$=86.3 (previous library \cite{RGW06}) to $F_0$=85.8 (modified library). The quantities $<$$F_i$$>$ are the 'mean overproduction factor' calculated from the mass of the respective isotope $i$ in the $p$-process zone divided by the total mass of the $p$-process zone times the respective solar mass fraction of the isotope. In Fig.~\ref{opf} our results with the modified reaction library are compared to the results published in \cite{RGW06} with the previous set of reaction rates to examine the influence of the newly implemented neutron capture data of KADoNiS v0.2. The result is a decrease in the normalized overproduction factor for almost all $p$ isotopes by an average value of -7\% (shown in the right part of Fig.~\ref{opf} as dashed line). The largest deviations occur for $^{84}$Sr (-20.5\%), $^{136}$Ce (+30.6\%), $^{156}$Dy (-39.2\%), $^{152}$Gd and $^{158}$Dy (-22.3\% each), $^{180}$W \mbox{(-22.9\%)}, and $^{190}$Pt (-32\%). 

In general, these differences are relatively strong in the mass region $A$=150-170. Reaction flux plots of this mass region reveal that the main flux proceeds by $(\gamma,\alpha)$, $(\gamma,n)$, $(n,\gamma)$, and $(n,\alpha)$ reactions, and is lower by 1-2 orders of magnitude compared to the previous reaction library. This drastic change cannot be explained by the implementation of larger experimental $(n,\gamma)$ cross sections of some of these $p$ isotopes, since this would in turn also increase the production channel via $(\gamma,n)$ reactions from heavier isotopes \cite{ID06}. Surprising at first glance, the cause of the differences are changes in the neutron rates at higher masses. A significant fraction of the seed abundances for the $p$ process is located in $^{209}$Bi and the Pb isotopes and converted to nuclei at lower mass by photodisintegration sequences starting with $(\gamma,n)$ reactions.

The importance of experimental data is strongly emphasized by these findings. Because of the magicity or near-magicity of the Pb and Bi isotopes, individual resonances determine the cross sections and Hauser-Feshbach theory is not applicable \cite{LL60,RTK96}. Furthermore, from discrepancies between resonance and activation measurements \cite{MHW77,BCM97} and from theoretical considerations \cite{RBO98}, it has been previously found that a small direct capture component contributes to neutron capture on $^{208}$Pb \cite{LL60,RBO98}. The interplay of resonant and direct capture contributions is difficult to handle in theoretical models and experiments prove to be indispensable. This also explains why there are deviations up to factor of 1.7-2 between the NON-SMOKER predictions and experimental data for these nuclei \cite{ID06}. In fact, Hauser-Feshbach models cannot be applied there. For the cases where the statistical model can be applied, the average NON-SMOKER uncertainty for stellar $(n,\gamma)$ cross sections is only $\pm$30\% or even better. 

The nuclides $^{152}$Gd and $^{164}$Er are not produced in the present simulation based on a 25 M$_\odot$ star. This is consistent with
previous work finding large $s$-process contributions to these nuclei. Also, the two odd-$A$ isotopes $^{113}$In and $^{115}$Sn are not produced. The latter underproduction problem is known since a long time \cite{ray90,raar95}. The initial seed abundances of $^{113}$In and $^{115}$Sn are destroyed by the $\gamma$ process, since the destruction channel is much stronger than the production channel. Thus, it appears as if the nuclides $^{152}$Gd, $^{164}$Er, $^{113}$In, and $^{115}$Sn have strong contributions from other processes and it is conceivable that they even may not be assigned to the group of $p$ nuclei.

Nemeth et al. \cite{NKT94} determined the contributions for $^{113}$In, and $^{115}$Sn with the (out-dated) classical $s$-process approach to be very small (less than 1\% for both cases). These calculations in the Cd-In-Sn region are complicated since many isomeric states have to be considered, and the $r$-process may contribute to the abundances of $^{113}$Cd and $^{115}$In. Although these two isotopes have quasi-stable ground-states, the $\beta$-decays of $r$-process progenitor nuclei can proceed via isomeric states: $^{113g}Ag$ $\rightarrow$ $^{113m}Cd$ $\rightarrow$ $^{113}$In and $^{115g}$Cd $\rightarrow$$^{115m}In$ $\rightarrow$ $^{115}$Sn. In \cite{NKT94} only part of the missing abundances could be ascribed to post-$r$-process $\beta$-decay chains, leaving rather large residues for other production mechanisms. In view of the progress achieved in $s$-process calculations using the TP-AGB star model, and with the availability of an improved set of $(n,\gamma)$ cross sections it appears worth while to update these older calculations. The new reaction library KADoNiS \cite{kado06} includes the latest Maxwellian averaged cross sections from very accurate time-of-flight measurements for the Cd and Sn isotopes \cite{WVT96,WVK96,KHW01,WVK01} and will soon be complemented by a measurement of the partial neutron capture cross section to the 14.1~y isomeric state in the $s$-process branching isotope $^{113}$Cd with the activation technique, which is underway at Forschungszentrum Karlsruhe. With this additional information new calculations are expected to provide more accurate $s$- and $r$-contributions for $^{113}$Cd and $^{115}$In. Based on these results, the $p$-contributions can be estimated as the residual via $N_p$=$N_\odot$--$N_s$--$N_r$.

\ack
This work was supported by the Swiss National Science Foundation
Grants 200020-061031 and 20002020-105328.


\end{document}